\documentclass[twocolumn]{article}
\usepackage{epsfig}
\usepackage{latexsym}
\usepackage{amsmath}
\usepackage{graphics}
\def\br{{\bf r}}
\def\bk{{\bf k}}

\begin{document}
\twocolumn[
\title{Rotons in gaseous Bose-Einstein condensates irradiated by a laser}
\author{D.H.J. O'Dell$^1$,
S. Giovanazzi$^{2}$ and G. Kurizki$^{3}$ \\
$^{1}$Physics and Astronomy Department,
University of Sussex, \\
Brighton BN1 9QH, UK \\
$^2$Center for Theoretical Physics and College of Science,
Polish Academy of Sciences \\
Aleja Lotnik\'ow, 32/46, 02-668 Warsaw, Poland \\
$^3$Chemical Physics Department, Weizmann Institute of Science, \\
76100 Rehovot, Israel }
\maketitle

A gaseous Bose-Einstein condensate (BEC) irradiated by a far
off-resonance laser has long-range interatomic correlations caused
by laser-induced dipole-dipole interactions. These correlations,
which are tunable via the laser intensity and frequency, can
produce a `roton' minimum in the excitation spectrum---behavior
reminiscent of the strongly correlated superfluid liquid helium
II.

According to the celebrated Bijl-Feynman formula \cite{feenberg69}
for the excitation spectrum of helium II
\begin{equation}
E(k) \le \frac{\hbar^{2} k^{2}}{2 m S(k)} \ , \label{eq:feynman}
\end{equation}
the peculiar ``roton'' minimum at $k \approx 2 \pi/r_{0}$, where
$r_{0}$ is the average atomic separation, is due to a
corresponding peak in the static structure factor $S(k)\equiv
\langle 0 \vert \hat{\rho}_{k} \hat{\rho}_{k}^{\dag} \vert 0
\rangle/N$. Here $N$ is the number of atoms of mass $m$, $\vert 0
\rangle $ the ground state of the system, and $\hat{\rho}_{k}
\equiv \sum_{q} \hat{c}^{\dag}_{q} \hat{c}_{q+k}$ the density
fluctuation operator. $S(k)$ is the Fourier transform of the pair
correlation function and hence provides a measure of the degree of
pair (2nd order) correlation between the atoms. The existence of
strong pair correlations in helium II may at first seem surprising
since it remains a liquid even at temperatures approaching
absolute zero precisely because of weak interatomic interactions
(in combination with a small atomic mass) \cite{huang}. However,
despite their apparent weakness these interactions are very
effective because the density of the liquid state is such that the
average atomic separation, $r_{0}=4.44$\AA, is close to the
minimum of the attractive interatomic potential well at 3\AA.

Contrast this now with an ultra-cold alkali atom gas in which the
Bose-Einstein condensed fraction can be very nearly 100 \%
\cite{nobel}. The interactions in $^{87}$Rb, for example, are
repulsive and characterized by an s-wave scattering length, $a
\approx 5.5$nm. This is between one and two orders of magnitude
smaller than the average atomic spacing at typical densities.
Steinhauer \textit{et al} \cite{steinhauer02} recently measured
the bulk excitation spectrum of a $^{87}$Rb BEC and found
excellent agreement with Bogoliubov theory \cite{llstatphys2}
(appropriate for a degenerate almost ideal Bose gas). There was no
roton minimum, a consequence of the diluteness with respect to
$a$. Indeed, since Eq.\ (\ref{eq:feynman}) becomes an equality
within the Bogoliubov theory \cite{feenberg69} one sees the pair
correlation is small compared to helium II. Significant pair
correlation might exist in gaseous BECs at the very small scale of
$a$, but this is fairly inaccessible in such a delicate system.

A marvellous feature of atoms though, is that their interactions
can be manipulated using external fields, allowing us to
microscopically engineer the macroscopic properties of a many-body
system. Thus the experiment of Inouye \textit{et al}
\cite{inouye98} took advantage of a Feshbach resonance to change
the s-wave scattering length using magnetic fields. We have
recently proposed the use of off-resonant lasers to induce
long-range dipole-dipole interactions whose characteristic length
is the laser wavelength. These interactions can cause laser
induced self-``gravity'' in a BEC, leading to 3-dimensional
self-trapping and electrostriction accompanied by unusual
excitation spectra \cite{odell2000}, as well as ``supersolid''
structures \cite{giovanazzi2002}. The aim here is to explore how
the excitation spectrum and, by virtue of (\ref{eq:feynman}), the
correlations of a gaseous BEC are changed when the interatomic
potential is modified via laser-induced dipole-dipole
interactions. This task requires a knowledge of the Fourier
transform (FT) of the total interatomic potential, so we turn to
this first.
\begin{figure}[htbp]
\begin{center}
\centerline{\epsfig{figure=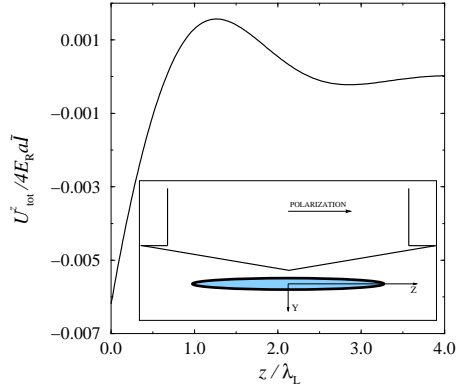,angle=-90,width=7cm}}
\end{center}
\caption{The total
(s-wave$+$dipole-dipole) 1D interatomic potential
$U^{z}_{\mathrm{tot}}(z)$ (FT of (\ref{eq:tot1dpot})), for
$w_{r}=3.5 \lambda_{\mathrm{L}}$. A repulsive contact term $4
E_{\mathrm{R}}  a (k_{\mathrm{L}} w_{r})^{-2}
(1+4\mathcal{I}/3)\delta(k_{\mathrm{L}}z)$ is not shown. Inset:
The laser beam and condensate geometry.} \label{fig:1dpot}
\end{figure}

Consider a BEC confined by a potential
$H_{\mathrm{trap}}=\frac{m}{2} \omega_{r}^2(x^2+y^2)+\frac{m}{2}
\omega_{z}^2z^2$ into a very elongated cigar shape ($\omega_{r}
\gg \omega_z$), irradiated by a far off-resonance plane-wave laser
(Fig.\ \ref{fig:1dpot}, inset). The laser polarization is along
the long z-axis of the condensate to suppress collective
(``superradiant'') Rayleigh scattering \cite{inouye99} or coherent
atomic recoil lasing \cite{piovella2001} that are forbidden along
the direction of polarization. The far off-resonance condition,
together with the small extent of the BEC along the laser
propagation direction, enables us to treat the electromagnetic
field inside the BEC in the Born approximation (field at each
point is the sum of the incident plus once-scattered fields). Then
the dipole-dipole potential between two atoms of separation ${\bf
r}$, induced by far-off resonance light of intensity $I$,
wave-vector ${\bf k}_{\mathrm{L}}= k_{\mathrm{L}} \hat{{\bf y}}$
(along the y-axis), and polarization $\hat{{\bf e}} = \hat{{\bf
z}}$ (along the z-axis) is \cite{thirunamachandran80}
\begin{equation}
U_{\mathrm{dd}}({\bf r}) = \frac{ I \alpha^{2}\left(\omega
\right)k_{\mathrm{L}}^3}{4 \pi c \varepsilon_{0}^{2}}
 V_{zz} \left(
k_{\mathrm{L}}, {\bf r} \right)
\cos \left(k_{\mathrm{L}} y \right).   \label{eq:tpot}
\end{equation}
Here $\alpha(\omega)$ is the isotropic, dynamic, polarizability of
the atoms at frequency $\omega=ck_{\mathrm{L}}= 2 \pi c/
\lambda_{\mathrm{L}}$. The pre-factor can be expressed in terms of
the Rayleigh scattering rate, $\gamma_{\mathrm{R}}$, as $I
\alpha^{2}k_{\mathrm{L}}^3/(4 \pi c \varepsilon_{0}^{2}) =(3/2)
\hbar \gamma_{\mathrm{R}}$. $V_{zz}$ is the component of the
retarded dipole-dipole interaction tensor generated by the
linearly $\hat{z}$-polarized laser light
\begin{eqnarray}
 V_{zz}  =  \frac{1}{k_{\mathrm{L}}^{3} r^{3}}  \Big[ \big(1 - 3
\cos^2 \theta \big) \big( \cos k_{\mathrm{L}} r + k_{\mathrm{L}}r
\sin
 k_{\mathrm{L}} r \big)  \nonumber \\
   -  \sin^2 \theta \ k_{\mathrm{L}}^{2}r^{2}
\cos  k_{\mathrm{L}}r \Big]
 \label{eq:retarded-dip-int}
\end{eqnarray}
$\theta$ being the angle between the  interatomic axis and the
z-axis. The far-zone ($ k_{\mathrm{L}}r \gg 1$) behavior of
(\ref{eq:tpot}) along the z-axis is proportional to $- \sin (
k_{\mathrm{L}}r) / ( k_{\mathrm{L}}r)^2$ and many atoms (400 at
densities of $8 \times 10^{14}$ atoms/cm$^3$) may lie within the
characteristic interaction volume ($\lambda_{\mathrm{L}}^{3}$) of
this attractive long-range potential. As for the electron gas and
charged Bose gas, mean-field (here Bogoliubov) theory applies in
this \emph{high density} regime \cite{foldy}.

The laser (dynamically) induced dipole-dipole potential is
distinguished from the static field ($r^{-3}$) case
\cite{goral,yi2001} by a longer range and a huge enhancement of
atomic polarizability around a resonance. For example, in
\cite{steinhauer02} $^{87}$Rb atoms are magnetically trapped in
the maximally stretched $\vert 5s  \hspace{1ex}^2S_{1/2},F=2,M=2
\rangle$ state. A laser polarized along $\hat{z}$ is then
$\pi$-polarized and only $\Delta M=0$ dipole transitions are
allowed. If the light is detuned by, say, $\delta=2 \pi
\times$(6.5GHz) (i.e.\ 1134 natural line widths) below the D1 line
(795.0nm) then only virtual transitions to the $\vert 5p
\hspace{1ex}^2P_{1/2},F=2,M=2 \rangle$ state need be considered.
We calculate $\alpha \approx 5.0 \times 10^{-35}
\mathrm{Cm}^{2}/\mathrm{V}$ (cf.\ the static value $5.3 \times
10^{-39} \mathrm{Cm}^{2}/\mathrm{V}$).

In terms of the condensate density $n(\mathbf{r})$ at zero
temperature, we account for atom-atom interactions using a
mean-field energy functional of the form $H_{\mathrm{dd}}
+H_{\mathrm{s}}$, where $H_{\mathrm{dd}} = (1/2) \int
n(\mathbf{r})
 \, U_{\mathrm{dd}}(\br-\br')\,  n(\mathbf{r}') \,
 d^{3}r \, d^{3}r'$, and $H_{\mathrm{s}}  =  (1/2) (4
\pi a \hbar^{2}/m) \int \, n(\mathbf{r})^{2} d^{3}r$ is due to
short-range interactions, which are described, as is usual, by a
delta function pseudo-potential (we take here the \emph{repulsive}
case for which $a>0$). By working with the bare dipole-dipole
interaction we assume the Born approximation also for
\emph{atom-atom} scattering by this long-range part of the total
potential. We note that the short-range (static) part of the
laser-induced dipole-dipole interaction can cause a shift in $a$.
For the laser intensities and detunings considered here this shift
is small according to existing estimates \cite{yi2001}.

In a radially tight trap it is reasonable to assume a
cylindrically symmetric ansatz for the density profile of radial
width $w_{r}$: $n(\br) \equiv N $ $(\pi w_{r}^{2} )^{-1}$ $
n^{z}(z) \exp \left[-(x^{2}+y^{2})/w_{r}^{2} \right] $, where N is
the total number of atoms and $n^{z}(z)$ is normalized to 1 and
left general. Denoting the FT of the atomic density by
$\tilde{n}(\bk)=\int \, \exp [-{\mathrm i} {\mathbf k} \cdot
{\mathbf r}] \, n({\mathbf r}) \, d^{3}r$, then we have
$H_{\mathrm{dd}}= (1/2)(2 \pi)^{-3}$ $ \int
\widetilde{U}_{\mathrm{dd}}({\mathbf k}) \, \tilde{n}(\bk) \,
\tilde{n}(-\bk) \, d^{3}k$, where the FT of the dipole-dipole
potential (\ref{eq:tpot}), $\widetilde{U}_{\mathrm{dd}}({\mathbf
k})= \int  \, \exp [-{\mathrm i} {\mathbf k} \cdot {\mathbf r}] \,
U_{\mathrm{dd}}({\mathbf r}) \, d^{3}r$, is the real part of
\begin{eqnarray}
\widetilde{U}_{\mathrm{dd}}({\mathbf k})  = \frac{{\mathrm I} \,
\alpha^{2}} {2 \;  \epsilon_{0}^{2} c} \Bigg(
\frac{k_{z}^2-k_{\mathrm{L}}^2}{k_{x}^{2}
+(k_{y}-k_{\mathrm{L}})^{2}+k_{z}^{2}-k_{\mathrm{L}}^2 -
\mathrm{i} \eta} \nonumber \\ +
 \frac{k_{z}^2-k_{\mathrm{L}}^2}{k_{x}^{2}
+(k_{y}+k_{\mathrm{L}})^{2}+k_{z}^{2}-k_{\mathrm{L}}^2 -
\mathrm{i} \eta} -\frac{2}{3} \Bigg) \;.
\label{eq:fourierdip-dippot}
\end{eqnarray}
 The principal value of the radial
integration in $H_{\mathrm{dd}}$ can be evaluated analytically so
that the dipole-dipole energy reduces to a one dimensional
functional along the axial direction $H_{\mathrm{dd}} = (N^{2}/2)
\int n^{z}(z)
 n^{z}(z') U^{z}_{\mathrm{dd}}(z-z') \, dz \, dz'$  $
 =   (N^{2}/4 \pi) \int \widetilde{n^{z}}(k_{z})
\widetilde{n^{z}}(-k_{z}) \widetilde{U^{z}_{\mathrm{dd}}}(k_{z})
\, d k_{z}$, where $\widetilde{n^{z}}(k_{z})$ is the FT
 of the axial density $n^{z}(z)$. The one-dimensional
(1D) axial potential that appears in this expression has the form
\begin{eqnarray}
\widetilde{U^{z}_{\mathrm{dd}}}(k_{z}) = \frac{{\mathrm I} \,
\alpha^{2} k_{\mathrm{L}}^2}{4 \pi \epsilon_{0}^{2} \; c} \,
Q(w_{r}, k_{z}) \quad
, \quad  \quad \nonumber \\
Q(w_{r}, k_{z}) = -\frac{2}{3}\frac{1}{k_{\mathrm{L}}^2 w_{r}^{2}}
 +   \frac{k_{z}^{2}-k_{\mathrm{L}}^{2}}{k_{\mathrm{L}}^2}
\mathrm{e}^{(k_{z}^{2}-2k_{\mathrm{L}}^{2} )
w_{r}^{2}/2} \times   \nonumber \\
 \quad  \sum_{j=0}^{\infty}
\frac{ (k_{\mathrm{L}} w_{r} )^{2j}}{2^{j} j {!}} \Re \left\{
E_{j+1} \left[\frac{(k_{z}^{2}-k_{\mathrm{L}}^{2} ) w_{r}^{2}}{2}
\right] \right\} \label{ueff1}
\end{eqnarray}
where $ \Re \{ E_{j}[z] \} $ is the real part of the generalized
exponential integral \cite{a+s}. The FT of the total (s-wave plus
dipole-dipole) 1D reduced interatomic potential is
\begin{equation}
\widetilde{U^{z}_{\mathrm{tot}}}(k_{z}) =4 E_{\mathrm{R}}
 a \left((k_{\mathrm{L}} w_{r})^{-2} +
\mathcal{I}Q(w_{r}, k_{z}) \right) \label{eq:tot1dpot}
\end{equation}
where $E_{\mathrm{R}}= \hbar^{2} k_{\mathrm{L}}^{2}/ 2m$ is the
photon recoil energy of an atom and $\mathcal{I}$ is the
dimensionless `intensity' parameter
\begin{equation}
{\mathcal I}= \frac{{\mathrm I} \, \alpha^{2}(\omega) m}{8 \pi
\epsilon_{0}^{2} c \hbar^{2} a}. \label{eq:itildedefn}
\end{equation}
It is emphasized that the radial degree of freedom is contained in
(\ref{eq:tot1dpot}) via the radius $w_{r}$. The coordinate space
potential, $U^{z}_{\mathrm{tot}}(z)$, is shown in Fig.\
\ref{fig:1dpot}.

We now have the essential ingredients to compute the excitation
spectrum of the BEC as it is the FT of the effective interatomic
interaction potential that appears in the Bogoliubov dispersion
formula \cite{llstatphys2}. Since the influence of radial
excitations upon the low-energy spectrum can be largely frozen out
under tight radial confinement, we shall consider only axial
phononic excitations and assume that the system is infinite along
this $\hat{z}$ direction. In terms of the phonon momentum
$p_{z}=\hbar k_{z}$, the axial Bogoliubov spectrum is (cf. Eq.\
(\ref{eq:feynman}))
\begin{equation}
E_{\mathrm{B}}=
  \sqrt{  c_{z}^{2}
p_{z}^{2}  + \left(p_{z}^{2}/2m \right)^2} = p_{z}^{2}/\left(2m S
\left( k_{z} \right) \right) \label{eq:scaledbog}
\end{equation}
where $c_{z}^{2}= \pi n(0)
w_{r}^{2}\widetilde{U^{z}_{\mathrm{tot}}}(k_{z})/m$. $n(0)$ is the
central density in the cigar, so $ \pi  n(0) w_{r}^{2} $ is the
linear density along the cigar. For the linear parts of the
spectrum, $c_{z}$ can be interpreted as the speed of sound in the
gas.
\begin{figure}[htbp]
\begin{center}
\centerline{\epsfig{figure=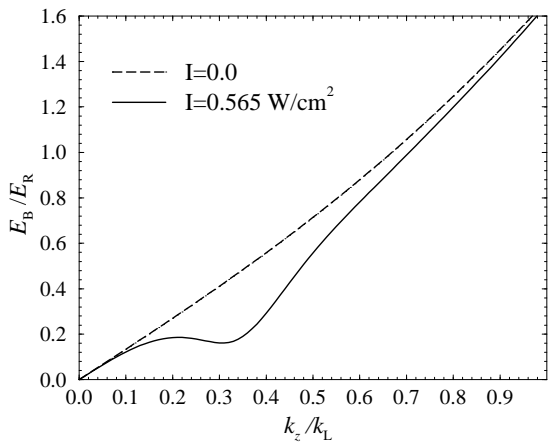,width=7cm,angle=-90}}
\end{center}
\caption{The Bogoliubov
dispersion relation for $^{87}$Rb with $w_{r}=3.5
\lambda_{\mathrm{L}}$ and $n(0)=8 \times 10^{20}$ atoms/m$^3$. For
pure s-wave scattering ($\mathcal{I}=0$) the inverse healing
length $1/\xi_{0}=\sqrt{8 \pi a n(0)} =1.32 k_{\mathrm{L}}$.} 
\label{fig:roton}
\end{figure}
Shining a 795.0nm laser upon a $^{87}$Rb BEC of density $n(0)=8
\times 10^{20}$atoms/m$^3$ and radius $w_{r}=3.5
\lambda_{\mathrm{L}}=2.78\mu$m, a `roton' minimum appears  when
$\mathcal{I} \geq 0.051$ (i.e.\ $I \ge 0.506$W/cm$^{2}$), although
the dispersion relation is considerably altered far before this.
The change in the dispersion relation could be observed using
Bragg spectroscopy as performed in \cite{steinhauer02}. Fig.\
\ref{fig:roton} plots the Bogoliubov dispersion for
$\mathcal{I}=0.057$ ($I = 0.565$W/cm$^{2}$).

\begin{figure}[htbp]
\begin{center}
\centerline{\epsfig{figure=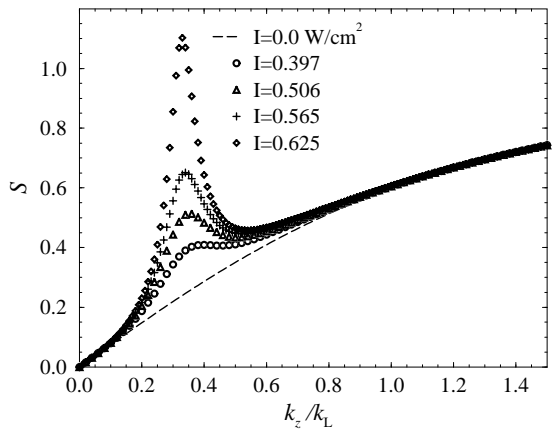,
  width=7cm,angle=-90}}
\end{center}
\caption{The static structure
factor, $S(k_{z})$ for various laser intensities. Same paramaters
as Fig.\ \ref{fig:roton}.} \label{fig:staticstruc}
\end{figure}

 Local to the `roton' minimum at
$k=k_{\mathrm{roton}}$ one can write $E = \Delta + \hbar^{2}
(k-k_{\mathrm{roton}})^{2}/ 2 m^{\ast}$, and for the parameters
above with $\mathcal{I}=0.057$ one finds $m^{\ast}=0.06m$. He II
has $m^{\ast}=0.16m$ \cite{llstatphys2}. The static structure
factor is plotted in Fig.\ \ref{fig:staticstruc}. The peak in
$S(k_{z})$ corresponds to the minimum in the energy spectrum. The
model described here predicts that when $\mathcal{I} \geq 0.066$
($I \geq 0.654$W/cm$^{2}$) the minimum touches the zero energy
axis. At this point the system is unstable to a periodic,
supersolid-like, density modulation
\cite{giovanazzi2002,pitaevskii84}.

The laser induced dipole-dipole potential can lead to
electrostriction (compression) of a condensate \cite{odell2000}.
In the present regime of low laser intensity/large detuning the
electrostriction is negligible (on a scale set by the collapse
threshold $\mathcal{I}=3/2$ \cite{odell2000,goral}). This regime
also ensures the absence of two-body bound states in the 1D
reduced potential shown in Fig.\ \ref{fig:1dpot}, a necessary
condition for the validity of the Born approximation for atom-atom
scattering by this potential. Only when $\mathcal{I} > 1.3$ do
bound states appear.

 The
interaction (\ref{eq:retarded-dip-int}) arises from the forward
scattering of laser photons by atom pairs. At large detunings
there are two main competing processes that can heat a dense gas :
A) \textbf{Light-induced transfer of pairs of colliding atoms to a
quasi-molecular excited} state followed by dissociation, releasing
$\approx \hbar \delta$ into the kinetic energy of the pair
\cite{burnett96}. This is a density dependent effect whose rate
can therefore be high. Even when the laser is red-detuned from an
atomic resonance, when two atoms collide the energy separation
between the ground state and a molecular excited state
($-C_{3}/r^{3}$) comes into resonance at small distances. However,
by choosing $\delta$ so that the resonance point occurs between
two molecular vibrational states this process is suppressed
\cite{burnett96}. Below the D1 line there are only discrete
molecular vibrational states (i.e. no continuum states) so a
detuning can be selected which is between these molecular
resonances \cite{miller93}, which are narrow at ultra-cold
temperatures. \newline B) \textbf{Incoherent light scattering by
single atoms} occurs at approximately the Rayleigh scattering rate
which can be written $\gamma_{\mathrm{R}} = (8/3) E_{\mathrm{R}}
k_{\mathrm{L}} a \mathcal{I}/\hbar$. Applying the f-sum rule for
the dynamic structure factor one can show \cite{pines+nozieres}
that Rayleigh scattering transfers energy to the gas at a rate
$\frac{d}{dt}E_{\mathrm{tot}}=2 E_{\mathrm{R}} N
\gamma_{\mathrm{R}}$ which, surprisingly, is independent of the
interactions between the atoms. Comparing this heating rate with
the energy of the ground state of the gas, $E_{\mathrm{tot}}
\approx H_{\mathrm{s}}+H_{\mathrm{trap}}+H_{\mathrm{kin}}$, where
$H_{\mathrm{kin}}$ is the kinetic energy of the atoms, one can
estimate a heating time via
$\tau_{\mathrm{heat}}=E_{\mathrm{tot}}/ (dE_{\mathrm{tot}}/dt)$.
To measure a roton the BEC must survive for longer than the roton
period $\tau_{\mathrm{roton}} \propto 2 \pi \hbar/ E_{R}$ (cf.
Fig.\ \ref{fig:roton}). For the density and radius stated above,
then for $\mathcal{I}=0.051$ one finds
$\tau_{\mathrm{heat}}\approx 8 \tau_{\mathrm{roton}}$, making an
experiment challenging but feasible.  The situation improves for
larger, denser, condensates since the polarization increases as
$N^2$, whereas $\gamma_{\mathrm{R}}$ is a single atom effect.
Finally, we note that reducing the s-wave scattering length via a
Feshbach resonance allows the laser intensity (and hence the
Rayleigh scattering) to be reduced by an equal factor---see Eq.\
(\ref{eq:itildedefn})---and still obtain the same effects.

In conclusion, atom-atom correlations due to laser induced
dipole-dipole interactions in a gaseous condensate can give a
roton minimum in the Bogoliubov dispersion relation. The
correlations are tunable via parameters such as radial width,
laser intensity and wavelength. We thank M. Boshier, C. Eberlein,
J. Steinhauer, R. Shiell, and H.T.C. Stoof for illuminating
discussions, and the Engineering and Physical Sciences Research
Council (EPSRC), the German-Israeli Foundation (GIF), and the EU
QUACS and CQG networks for funding.





\end{document}